\documentclass[runningheads]{llncs}
\usepackage{graphicx}
\usepackage{tikz}
\usetikzlibrary{shapes}
\usepackage{fontawesome}

\usepackage{listings}
\usepackage{color}
\usepackage{courier}

\begin{document}
\title{Using the Shapes Constraint Language for modelling regulatory requirements}
\titlerunning{Using SHACL for modelling regulatory requirements}
% If the paper title is too long for the running head, you can set
% an abbreviated paper title here
%
\author{Veronika Heimsbakk\inst{1} \and Kristian Torkelsen\inst{2}}
\authorrunning{V. Heimsbakk, K. Torkelsen}
% First names are abbreviated in the running head.
% If there are more than two authors, 'et al.' is used.
%
\institute{Capgemini Norge AS\\ \email{veronika.heimsbakk@capgemini.com} \and
Norwegian Maritime Authority\\
\email{krt@sdir.no}}
\maketitle              % typeset the header of the contribution
\begin{abstract}
Ontologies are traditionally expressed in the Web Ontology Language (OWL), that provides a syntax for expressing taxonomies with axioms regulating class membership. The semantics of OWL, based on Description Logic (DL), allows for the use of automated reasoning to check the consistency of ontologies, perform classification, and to answer DL queries. However, the open world assumption of OWL, along with  limitations in its expressiveness, makes OWL less suitable for modelling rules and regulations, used in public administration. In such cases, it is desirable to have closed world semantics and a rule-based engine to check compliance with regulations. In this paper we describe and discuss data model management using the Shapes Constraint Language (SHACL), for concept modelling of concrete requirements in regulation documents within the public sector. We show how complex regulations, often containing a number of alternative requirements, can be expressed as constraints, and the utility of SHACL engines in verification of instance data against the SHACL model. We discuss benefits of modelling with SHACL, compared to OWL, and demonstrate the maintainability of the SHACL model by domain experts without prior knowledge of ontology management.

\end{abstract}
%
%
%

%% INTRODUCTION
\section{Introduction}

Supervisory systems play a key role in activities carried out by public administrations. At the Norwegian Maritime Authority (NMA) a replacement of their supervision system was needed. The supervision system determines whether a certain vessel or person comply with a set of requirements across mandatory maritime statues and regulations. Compliance with relevant requirements is necessary before applicable certificates are issued to ships or persons who apply for them.

The requirements are concrete and are considered unambiguous in their interpretation. When modelling regulatory requirements for the new supervision system in detail, it became apparent that mapping information into RDF Schema (RDFS) \cite{rdfs} and OWL \cite{owl2} had limitations. One of NMA's requirements for the supervision system was to automatically detect what a specific individual was missing for a given certificate. Towards this, describing information under a closed world assumption was needed. Regulatory requirements adopted by the NMA often contain complex permutations of alternative requirements equally valid to one another. This required a modelling expressiveness that RDFS and OWL could not provide in a manner that could be maintained by non-ontologists. This missing expressiveness can be fulfilled using the W3C recommended Shapes Constraint Language \cite{shacl} which, as this paper illustrates, gave us promising results regarding our modelling issues. 

\section{Background}
\label{sec:background}

\subsection{Norwegian Maritime Authority}
The Norwegian Maritime Authority is a public administration organ with legal powers over ships flying the Norwegian flag and all vessels operating in waters under Norwegian jurisdiction. The NMA is responsible for matters related to safety of life, health, environment and material values on ships \cite{sdir}. Its activities are governed by international and national legislation, political decisions, and agreements.
The purpose of NMA's supervision ensures that vessels are in good shape and are safe workplaces manned with qualified seafarers. NMA issues certificates to ships and shipping companies, and educational institutions. Certification, document control, inspection, and auditing to ensure compliance with the legislation are activities carried out by the NMA. 
In matters concerning ship and environmental safety, the NMA aims to be a driving force for the shipping industry, and national and international policy makers.

\subsection{NMA supervision system}
NMA's current supervision system is a legacy system that wherein updating and adding new data is time consuming, intricate and costly. Data is spread across unconnected silos. The user interface is poor and making maintenance and management difficult for the domain experts. The purpose of the new solution is to facilitate interaction between all stakeholders that are involved in ensuring that a vessel is compliant with relevant regulations.

Inspired inter alia by the implementations at the Norwegian Broadcasting Corporation \cite{nrkorigo}, which is building a metadata bank using semantic web technologies, the NMA chose to build their new solution based upon RDF.

\section{Solution architecture}
\label{sec:arch}

The source of data comprises 150 documents, mostly regulations, maintained by the NMA. The solution architecture required that all requirement descriptions, legal scopes and concepts be semantically modelled using RDF.

Initial attempts to model the transformation to graph manually were found to be time consuming and inefficient. Both from budget and schedule perspectives. Estimated time on manual modelling was 12 000 working hours. Moreover, manual modelling increases the risk of overseeing important information and misinterpretation of concepts. Especially so when performed by ontologists without domain expertise. 

\begin{figure}
\centering
\scalebox{0.75}{
\begin{tikzpicture}
\node[cylinder, draw=black, rotate=90, minimum width=1cm, minimum height=1cm] (text) at (0.25,0) {};
\node[] (text) at (0.25,0) {\faFileTextO};
\node[align=center] (x) at (0.25,-0.7) {text service};
\node[rectangle, draw=black, minimum width=1.5cm, minimum height=1.5cm, align=center] (nlp) at (3,0) {NLP\\module};
\node[rectangle, draw=black, minimum width=1.5cm, minimum height=1.5cm, align=center] (rdf) at (7,0) {RDF\\transformation};
\node[cylinder, draw=black, rotate=90, minimum width=1.5cm, minimum height=1.2cm] (train) at (3,-2) {};
\node[align=center] (x) at (3,-2) {training\\sets};
\node[rectangle, draw=black, minimum width=1.2cm, minimum height=1cm] (ottr-s) at (7.1,2.1) {};
\node[rectangle, fill=white, draw=black, minimum width=1.2cm, minimum height=1cm] (ottr) at (7,2) {\scriptsize{OTTR}};
\node[rectangle, draw=black, minimum width=2.4cm, minimum height=1cm] (ano-s) at (3.1,-3.9) {};
\node[rectangle, fill=white, draw=black, minimum width=2.4cm, minimum height=1cm] (ano) at (3,-4) {Annotations};
\node[rectangle, draw=black, minimum width=3cm, minimum height=3.5cm] at (11,0) {};
\node[rectangle, draw=black, minimum width=1cm, minimum height=3cm] at (10.25,0) {};
\node[rectangle, draw=black, minimum width=0.8cm, minimum height=0.5cm] (dev) at (10.25,0) {\scriptsize{dev}};
\node[rectangle, draw=black, minimum width=0.8cm, minimum height=0.5cm] (prod) at (10.25,-1) {\scriptsize{prod}};
\node[cylinder, draw=black, rotate=90, minimum width=1cm, minimum height=1.2cm] (db) at (11.6,-1) {};
\node[align=center] (sparql) at (13.5,-0.75) {SPARQL};
\node[align=center] (graphql) at (13.5,-1.25) {GraphQL};
\node[align=center] at (10.25,1) {RDF};
\node[align=center] at (11.55,1) {Triple\\store};

\node[rectangle, draw=black] (t) at (8,-2) {ttl-to-csv};

\draw[->, thick] (0.75,0)--(nlp);
\draw[->, thick] (train)--(nlp);
\draw[->, thick] (ano)--(train);
\draw[->, thick] (nlp)--node[above]{JSON} (rdf);
\draw[->, thick] (ottr)--(rdf);
\draw[->, thick] (rdf)--node[above]{API}(dev);
\draw[->, thick] (dev)--(prod);
\draw[->, thick] (prod)--(db);
\draw[->, thick] (sparql)--(11.75,-0.75);
\draw[->, thick] (graphql)--(11.75,-1.25);
\draw[thick] (prod) -- (10.25,-2) -- (t);
\draw[->, thick] (t)--node[above]{API}(train);
\end{tikzpicture}
}
\caption{Project pipeline of the semantic modules.}
\label{fig:pipeline}
\end{figure}
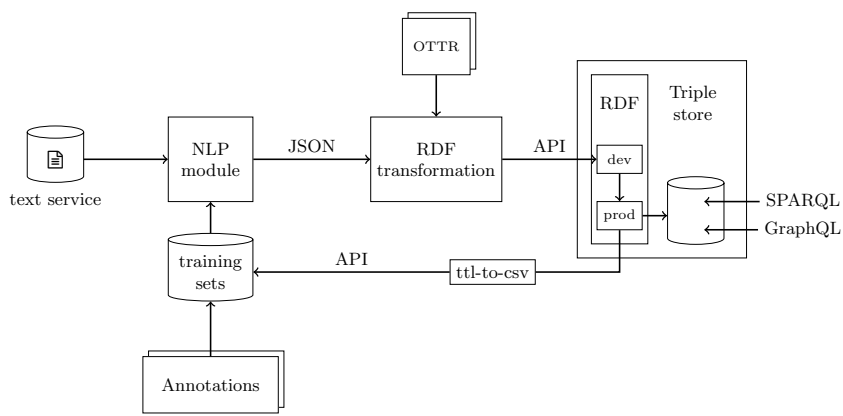

The solution introduced extracting context, concepts and relationships from the regulations using Natural Language Processing (NLP) techniques \footnote{\url{https://github.com/Sjofartsdirektoratet/NLP\_PoC}}. Using pattern matching with spaCy\footnote{\url{https://spacy.io/}} and Named Entity Recognition (NER) for entity classification, gave expected results. The NER-model is trained using annotations in random excerpts of a subset of regulations. Annotations are performed by seven maritime domain experts at the NMA using prodigy\footnote{\url{https://prodi.gy/}}. The final implementation are able to automatically generate semantic knowledge graphs directly from plain text. Fig. \ref{fig:pipeline} displays an overview of the semantic modules in our project pipeline. By introducing this pipeline, the cost of knowledge modelling was reduced by 10 000 working hours, compared to the original estimate.

Transforming the result of the NLP service into RDF is implemented in Java using rdf4j \footnote{\url{https://rdf4j.org/}}, in combination with Reasonable Ontology Templates (OTTR) \cite{ottr} for RDF serialization. The resulting graph contains SHACL shapes describing regulatory requirements, legal scopes, and the relationships between them. In addition, a lightweight OWL ontology are created based on NER model results. The main purpose of our OWL ontology is to keep track of the common vocabulary and assist in the work of maintenance of regulations, and authoring new regulations. The OWL ontology can then be returned to the NER-model to continue training the model in order to achieve higher accuracy in classification.

%% DESCRIBING REQUIREMENTS
\section{Modelling NMA regulations}
\label{sec:description}

RDF is a simple and easy to understand data format when expressed in human-readable syntax as RDF Turtle \cite{turtle}. Being able to read complex constraints requires minimal training for domain experts at the NMA, and is easier for actors at the NMA to read compared to models in OWL with Turtle.

This section will introduce you to a specific example of a regulatory requirement modelled in both OWL and SHACL for comparison.

\begin{figure}[h!]
\emph{\textbf{§28 2)} In order to be issued a competency certificate deck officer class 1, in addition to the requirements in § 23, § 24, § 25, § 26 and § 27, a minimum of 36 months seagoing service as a responsible deck officer on seagoing ships with a gross tonnage of more than 500 is required. Seagoing service is reduced to 24 months if at least 12 months are earned as chief officer on a vessel of gross tonnage 500 or more.}
\caption{Paragraph 28, part 2 from regulation no. 1523, describing a competency certificate for sailors.}
\label{fig:par28}
\end{figure}

As the excerpt in Fig. \ref{fig:par28} shows, the amount of important information is considerable and so is its complexity. Concepts, references, relationships, and data values make up the building blocks of the data model. This particular excerpt also describes two different alternatives that are equally valid.

\begin{itemize}
\item A minimum 36 months as a deck officer; or
\item A minimum 24 months as a deck officer, where at least 12 months is served as a chief officer.
\end{itemize}

Here, "deck officer" is a superclass for all deck officer positions, and "chief officer" is a subclass of the class "deck officer". In addition, there is additional information connected to these alternatives, regarding specification of the vessel and seagoing service in question.

\subsection{Web Ontology Language}
Ontologies are usually expressed using the Web Ontology Language, that provides a syntax for expressing taxonomies with axioms regulating class membership, groups of things, and relationships between things. Knowledge expressed in OWL can apply automated reasoning to an ontology in order to check consistency of ontologies, perform classification, and to answer DL queries \cite{owl2}.

However, the open world assumption of OWL makes it less suitable for modelling rules and regulations. Exploiting the reasoning capabilities of OWL is useful for inferring new facts and discover inconsistency. However, a reasoner is unable to detect missing focus nodes \cite{knublauch1}. Being able to describe a data model that does not allow incomplete knowledge is necessary for a semantic knowledge graph describing regulations. The open world assumption assumes that a statement is true, whether or not it is known to be true. However, in a regulatory domain the assumption of that everything that is not currently known to be true is false is applicable.

If we were to model regulatory requirements using the Web Ontology Language, we would have to model OWL axioms regulating class membership. In order to understand OWL axioms and consequences of OWL semantics, understanding of discrete mathematics and logic is required. As the solution is to be maintained and managed by maritime domain experts at the NMA, with limited experience in semantic technologies, a solution with high readability was required. Since there are no existing open source technologies which manage SHACL shapes through a user interface\footnote{\url{https://github.com/fekaputra/shacl-plugin}}, in the same way Protégé does for OWL, it is crucial to be able to read raw RDF files containing shapes. The data owner is an expert on the maritime domain, not on semantic technologies.

\subsubsection{Modelling regulatory requirements using OWL}

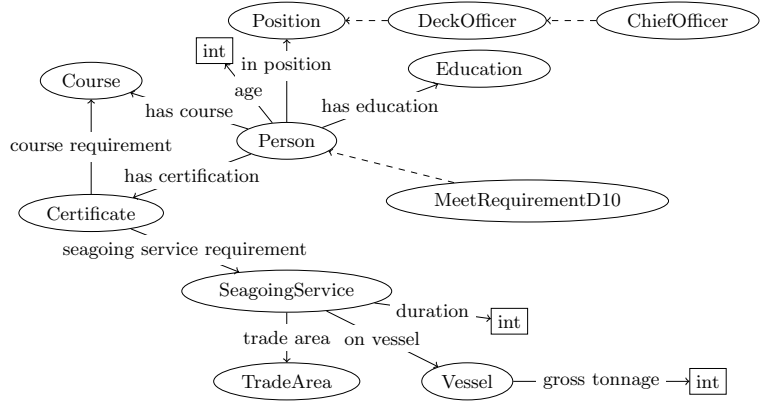
\begin{figure}[h!]
\centering
\scalebox{0.8}{
\begin{tikzpicture}
\node[ellipse, draw=black] (person) at (0,0) {Person};
\node[ellipse, draw=black] (position) at (0,2) {Position};
\node[ellipse, draw=black] (sgs) at (0,-2.5) {SeagoingService};
\node[ellipse, draw=black] (edu) at (3.2,1.2) {Education};
\node[ellipse, draw=black] (d10) at (4,-1) {MeetRequirementD10};
\node[ellipse, draw=black] (course) at (-3.25,1) {Course};
\node[ellipse, draw=black] (cert) at (-3.25,-1.2) {Certificate};
\node[ellipse, draw=black] (ta) at (0,-4) {TradeArea};
\node[ellipse, draw=black] (vessel) at (3,-4) {Vessel};
\node[rectangle, draw=black] (int1) at (-1.2,1.5) {int};
\node[rectangle, draw=black] (int2) at (7,-4) {int};
\node[rectangle, draw=black] (int3) at (3.7,-3) {int};
\node[ellipse, draw=black] (do) at (3,2) {DeckOfficer};
\node[ellipse, draw=black] (co) at (6.5,2) {ChiefOfficer};

\draw[->] (person) --node[fill=white] {age} (int1);
\draw[->] (person) --node[fill=white, above] {in position} (position);
\draw[->] (person) --node[fill=white] {has certification} (cert);
\draw[->] (person) --node[fill=white] {has course} (course);
\draw[->] (person) --node[fill=white] {has education} (edu);
\draw[->, dashed] (d10) --node[] {} (person);
\draw[->] (cert) --node[fill=white] {seagoing service requirement} (sgs);
\draw[->] (cert) --node[fill=white] {course requirement} (course);
\draw[->] (sgs) --node[fill=white] {trade area} (ta);
\draw[->] (sgs) --node[fill=white] {duration} (int3);
\draw[->] (sgs) --node[fill=white] {on vessel} (vessel);
\draw[->] (vessel) --node[fill=white] {gross tonnage} (int2);
\draw[->, dashed] (do) --node[] {} (position);
\draw[->, dashed] (co) --node[] {} (do);
\end{tikzpicture}
}
\caption{Visualization of the OWL class hierarchy and relationships used for the axioms in Fig. \ref{fig:owl}. Subclasses denoted using dashed lines.}
\label{fig:owlviz}
\end{figure}

In order to determine whether a person shall fulfil a specific set of requirements, we must define OWL restrictions onto a subclass of person. These restrictions describe the alternatives required for achieving the certificate in question. A proposed modelling solution in OWL is shown in Fig. \ref{fig:owl}, with class relations as visualized in Fig. \ref{fig:owlviz}. The OWL example is shown in OWL Manchester syntax due to readability of the example for the purpose of this paper. By performing reasoning with HermiT\footnote{http://www.hermit-reasoner.com/}, the reasoner could infer individuals matching either of the two alternatives described in Fig. \ref{fig:owl}. A reasoner will however only detect logical contradictions in our restrictions, not state what resources are missing in order to fulfil the requirements.

\begin{figure}
\begin{lstlisting}
(has seagoing service SOME 
	(SeagoingService 
	AND (on vessel SOME (Vessel 
		AND gross tonnage SOME xsd:int[>= 500]))
	AND (in position SOME DeckOfficer)
	AND (trade area VALUE TradeArea3)
	AND (duration SOME xsd:int[>= 1080]))
AND (age SOME xsd:int[>= 20])
AND (has course VALUE MedicalCourse)
AND (has certificate VALUE GMDSSRadioCertificate AND 
     has certificate VALUE HealthCertificate)
AND (has certificate VALUE D2A0 OR has certificate VALUE D2B0 OR
     has certificate VALUE D3A0 OR has certificate VALUE D3B0 OR
     has certificate VALUE D4B0 OR has certificate VALUE D4F0)
AND (has course VALUE VSK OR has course VALUE VSKR OR has course VALUE OGD))

OR

(
	(has seagoing service SOME 
		(SeagoingService
		AND (on vessel SOME (Vessel
			AND gross tonnage SOME xsd:int[>= 500]))
		AND (in position SOME DeckOfficer)
		AND (trade area VALUE TradeArea3)
		AND (duration SOME xsd:int[>= 720])))
	AND 
	(has seagoing service SOME 
		(SeagoingService 
		AND (on vessel SOME (Vessel
			AND gross tonnage SOME xsd:int[>= 500]))
		AND (in position SOME ChiefOfficer)
		AND (trade area VALUE TradeArea3)
		AND (duration SOME xsd:int[>= 360])))

	AND (age SOME xsd:int[>= 20])
	AND (has course VALUE MedicalCourse)
	AND (has certificate VALUE GMDSSRadioCertificate AND 
      has certificate VALUE HealthCertificate)
	AND (has certificate VALUE D2A0 OR has certificate VALUE D2B0 OR 
      has certificate VALUE D3A0 OR has certificate VALUE D3B0)
	AND (has course VALUE VSK OR has course VALUE VSKR OR has course VALUE OGD)
)
\end{lstlisting}
\caption{Modelling our regulatory example in OWL Manchester syntax.}
\label{fig:owl}
\end{figure}

\subsection{Shapes Constraint Language}

The Shapes Constraint Language are designed for validating RDF under a closed world assumption \cite{shacl}. It is therefore possible to detect missing statements in a person's instance data against a set of requirements.

OWL restrictions are not data constraints in the manner that they do not conform data to the restrictions given. Restrictions are rather steps in reasoning, inferences, applied based on the OWL restrictions. You cannot ask a reasoner to conform to a given set of constraints, as there are no concept of false conformance in OWL, hence the open world assumption. OWL, and RDFS, were designed for inferencing, not data validation.

In addition, SHACL is more expressive than the currently available alternatives \cite{knublauch2} \cite{validatingrdfdata}, as described in Sect. \ref{sec:compare}. This makes it possible to model requirements in great detail. 

\subsubsection{Shapes}
The Shapes Constraint Language are built up by shapes, which are constraints for given RDF resources \cite{shacl}. There are two kinds of shapes in SHACL.
\begin{enumerate}
\item \textit{Node shapes} describing constraints about focus nodes.
\item \textit{Property shapes} describing constraints about predicate and/or object value of a triple.
\end{enumerate}

A node shape is a shape that \textit{is not} the subject of a triple where \texttt{sh:path} is the predicate. While a property shape is a shape that \textit{is} the subject of a triple where \texttt{sh:path} is the predicate. We can add property shapes onto a node shape using the resource \texttt{sh:property}. 

\begin{figure}[h!]
\begin{verbatim}
:VesselShape
  a sh:NodeShape ;
  sh:targetClass :Vessel ;
  sh:property :GTShape .

:GTShape
  a sh:PropertyShape ;
  sh:path :grossTonnage .
\end{verbatim}
\caption{Minimal example of a node shape and property shape combined.}
\label{fig:nodeprop}
\end{figure}

Fig. \ref{fig:nodeprop} contains a node shape that targets the class \texttt{:Vessel}. This means that every constraint following in this shape is a constraint on instances of the class \texttt{:Vessel}. Further, it says that any instance of the class \texttt{:Vessel} may have some relationship \texttt{:grossTonnage} pointing to, in this case, some random value, as we do not have any more detailed constraints in the property shape.

\subsubsection{Constraints}
There are numerous constraints available in the SHACL Core Constraint vocabulary \cite{shacl}. SHACL is backed by SPARQL-queries, making the vocabulary self-descriptive and therefore extensible with custom constraints.

% Skrive noe om hva ontologien kan brukes til; resonnering på tvers av forskrifter

%% SHACL AT NMA
\subsection{Modelling SHACL at NMA}
\label{sec:nma}

Being able to model regulatory requirements as AND- and OR-constraints, as shown in Fig. \ref{fig:alt} in a human-readable syntax is a great benefit for domain experts at the NMA for the purpose of future graph maintenance. 

This section will explore our solution to modelling regulatory requirements at the NMA using the Shapes Constraint Language.

\subsubsection{Relationships}
\begin{figure}[h!]
\begin{verbatim}
:GT500
  a sh:PropertyShape ;
  sh:path :grossTonnage ;
  sh:minInclusive 500 ;
  sh:datatype unit:GT ;
  sh:minCount 1 ;
  sh:maxCount 1 .
\end{verbatim}
\caption{Property shape describing the gross tonnage relationship.}
\label{fig:gt}
\end{figure}

Fig. \ref{fig:gt} shows a constraint describing the relationship of "gross tonnage". Describing gross tonnage as a value of minimum or inclusive 500 of the datatype \texttt{unit:GT}, which is the Quantities, Units, Dimensions, and Types Ontology's \footnote{\url{http://qudt.org/vocab/unit/}} definition of gross tonnage. There is a min and max count of one, stating that this relationship is unique and mandatory for every occurrence of this predicate.

\subsubsection{Alternatives}
As shown in Fig. \ref{fig:par28}, there are alternatives described in the regulations. At the NMA, logical constraint components are used to model alternatives.

\begin{figure}[h!]
\begin{lstlisting}
...
sh:or (
  [ sh:and ( # first alternative
    [ sh:or (cert:PS_D2A0 cert:PS_D2B0 cert:PS_D3A0 
             cert:PS_D3B0 cert:PS_D4B0 cert:PS_D4F0) ]
    [ sh:path :hasSeagoingServiceRequirement ;
      sh:hasValue :SGS_500_1080_DO ; 
      sh:order 1 ; ]
  )]

  [ sh:and ( # second alternative
    [ sh:or (cert:PS_D2A0 cert:PS_D2B0 cert:PS_D3A0 cert:PS_D3B0) ]
    [ sh:path :hasSeagoingServiceRequirement ;
      sh:hasValue :SGS_500_720_DO ; 
      sh:order 2 ; ]
    [ sh:path :hasSeagoingServiceRequirement ;
      sh:hasValue :SGS_500_360_CO ; 
      sh:order 2 ; ]
  )]
) ;
...
\end{lstlisting}
\caption{Alternatives of a certificate described using logical constraint components.}
\label{fig:alt}
\end{figure}

Fig. \ref{fig:alt} shows the approach used to model alternatives for achieving a master mariner certificate. A SHACL OR-constraint is a list-taking constraint that, in this case, takes two AND-constraints as its list items. The first AND-constraint is the first alternative. It takes two items in its list:
\begin{enumerate}
\item An OR-list of certificates. At least one of these certificates must be present for the whole expression to be valid.
\item A specific value for a seagoing service requirement.
\end{enumerate}

The second AND-constraint is the second alternative. It takes three items in its list:
\begin{enumerate}
\item An OR-list of certificates.
\item A specific value for a seagoing service requirement.
\item Another specific value for a seagoing service requirement.
\end{enumerate}

\texttt{sh:order} is a non-validating resource used for distinguishing seagoing service alternatives. Seagoing service requirements are described as node shapes containing constraints connected to this particular instance of a seagoing service. In practise, this means that the model contains various permutations of possible seagoing services. The remaining constraints contains the rest of the information connected to the seagoing service, as described in Fig. \ref{fig:par28}. Apart from gross tonnage, as described in Fig. \ref{fig:gt}, there is information about duration and position. We also have the concept of a trade area, described in other paragraphs referenced from the one in Fig. \ref{fig:par28}.

\begin{figure}[h!]
\begin{verbatim}
:Duration1080
  a sh:PropertyShape ;
  sh:path :duration ;
  sh:minInclusive 1080 ;
  sh:datatype unit:DAY .

:PositionDO
  a sh:PropertyShape ;
  sh:path :inPosition ;
  sh:class :DeckOfficerPosition .

:TradeAreaBF
  a sh:PropertyShape ;
  sh:path :tradeArea ;
  sh:hasValue :BankFishing .
\end{verbatim}
\caption{Constraints describing first alternative of seagoing service in Fig. \ref{fig:alt}, in addition to property shape described in Fig. \ref{fig:gt}.}
\label{fig:constraints}
\end{figure}

\begin{figure}[h!]
\begin{verbatim}
:SeagoingServiceURI
  a sh:NodeShape ;
  sh:targetClass SGS_500_1080_DO ;
  sh:property :Duration1080, :PositionDO, :TradeAreaBF, :GT500 .
\end{verbatim}
\caption{Complete node shape for first alternative of seagoing service shown in Fig. \ref{fig:alt}.}
\label{fig:nodeshapecomplete}
\end{figure}

%% VALIDATION
\subsection{Exploiting the validation capabilities of SHACL}
\label{sec:validation}
By modelling certification requirements as SHACL shapes, one can exploit the validation capabilities of the language. The ability to identify specific requirements which a sailor does \textit{not} fulfil in order to achieve a certificate is a highly apparent implementation for the project's pipeline.

By running the shapes graph and instance graph through an SHACL engine, a validation report is returned as RDF, either conforming true or false. If the report conforms false, a validation result is produced, containing detailed information on inconformity of instance data. The validation report can be used by applications in order to make computational decisions on how to handle inconformity.

%% COMPARED
\section{SHACL compared to other validation languages}
\label{sec:compare}
\subsection{SPARQL Inferencing Notation (SPIN)}
The high-level languages of both SPIN and SHACL, the SPIN Templates and SHACL Constraint Components, are backed by SPARQL-queries that implement executable semantics. The vocabularies are therefore self-descriptive and extensible. Constraint components are however, more flexible than templates, due to the fact that it is possible to combine multiple constraint types into the same shape definition. SPIN Templates require new instances and multiple \texttt{spin:constraint} triples.

Comparing similarities of inferencing rules of both SPIN and SHACL is not relevant for the project at the NMA yet, and is therefore not evaluated in this paper.

SHACL includes equivalent features for every feature in SPIN, plus even more \cite{knublauch2}. SHACL is also an official W3C Recommendation \cite{shacl}, which is preferable (although not necessary) for a  project in production.

\subsection{Shape Expressions (ShEx)}
ShEx and SHACL share the same goal; to describe and validate RDF data using a high-level language \cite{validatingrdfdata}. Both languages contains the concept of shapes on RDF resources, both shapes on nodes and properties. Although both languages share a common goal, their design are based on different approaches.

ShEx intends to be a grammar or schema for RDF, while SHACL is aiming to provide a constraint language for RDF. The main goal of SHACL is to verify that a given RDF graph conforms to a collection of constraints \cite{shacl}. This difference is reflected in the validation results. ShEx returns an annotated graph containing the RDF graph that were validated \cite{validatingrdfdata}. SHACL however describes in detail the errors returned when the validation does not conform. 

\section{Discussion}
\label{sec:discussion}

RDF is an easy to understand data format when expressed in a human-readable syntax, like Turtle. It requires a few hours of training for non-semantic users to start writing plain RDF Turtle by hand. Modelling complex data and real-life problems, however, requires quite different skills than the ability of understanding RDF itself. The Web Ontology Language offer the ability to express abstract concepts as sets, applying restrictions and axioms on sets and their relationships. Being able to create such models require an understanding of set theory, relations, propositional logic and first-order logic. This increases complexity of future graph maintenance in projects where data owners are not experienced in the field of logic. By expressing the data model using the Shapes Constraint Language we are able to preserve the simplistic RDF Turtle structure. Which makes it easier for the NMA to perform future graph maintenance.

The Web Ontology Language is suitable for modelling an ontology for a common understanding of concepts and terminology of the maritime domain. Which is beneficial for authoring new regulations. However, it is less suitable for modelling regulatory requirements. This is due to the open world assumption of OWL. Regulatory requirements do not allow for incomplete knowledge. Therefore based on our experience we recommend the use of SHACL as a preferred language in the regulatory domain. Notably, SHACL allows for modelling knowledge under a closed world assumption, meaning that statements are either true or false. 

Modelling detailed and machine readable regulatory requirements using SHACL makes it possible to move computational logic from the project's back end to the graph itself. The validation capabilities of SHACL is suitable for identifying missing instance data in order to compare a sailors CV to a set of requirements related to a certificate. 

Working closely with domain experts at the NMA has been crucial for a thorough understanding of maritime data and their regulatory requirements. By having a close cooperation between ontologists and domain experts we have developed SHACL shapes that reflects requirements in great detail. That enables the full potential in SHACL validation and graph querying. Making information and knowledge available for project applications and business user to a great extent. 

%% MORE ON TIME SAVING AND EFFICIENCY CONCLUDING THE PROJECT ON BUDGET AND SCHEDULE

%% CONCLUSION
\section{Conclusion}
\label{sec:conclusion}

SHACL offers a rich vocabulary that allows us to model complex real-life problems under a closed world assumption. SHACL has proven itself to be more expressive than current alternatives as SPIN and ShEx. And relying on SPARQL, its vocabulary is also extensible. The closed world assumption of SHACL makes it suitable for applications not allowing for incomplete knowledge in their data, as is the case of regulatory data. 

One can identify missing focus nodes in instance data by performing validation against a set of constraints. This is beneficial for use cases wherein returning a set for missing statements to the end user is important. As also for identifying what experiences a sailor is missing in order to achieve a specific certificate.

However, there is a need for available modelling tools providing a graphical user interface in order to outreach the potential of SHACL, and RDF for that matter, as a data model language for the enterprise. We believe this will be beneficial for increasing the popularity of semantic knowledge graphs to a broader audience. 

\section*{Acknowledgements}
We thank \textbf{Christian Mahesh Hansen} (Aibel AS) for his input on OWL, \textbf{Aniruddha Khadkikar} (Capgemini), \textbf{Calum Chalmers} (Capgemini) and \textbf{Arild Viddal} (the Norwegian Maritime Authority) for thorough proofreading, and \textbf{Robert Engels} (Capgemini) for his unconditional support.

% ---- Bibliography ----
%
% BibTeX users should specify bibliography style 'splncs04'.
% References will then be sorted and formatted in the correct style.
%
% \bibliographystyle{splncs04}
% \bibliography{mybibliography}
%

\end{document}